\documentclass{elsart}

 \usepackage{graphicx}

\usepackage{amssymb}

\begin{document}

\begin{frontmatter}



\title{Transient differential reflectivity of ferromagnetic and paramagnetic phases in the bilayered manganite La$_{1.24}$Sr$_{1.76}$Mn$_2$O$_7$}


\author{Y. Hirobe\corauthref{cor1}},
\corauth[cor1]{Tel.: +81 332384108; fax: +81 332383430.}
\ead{y-hirobe@sophia.ac.jp}
\author{Y. Kubo},
\author{K. Kouyama},
\author{H. Kunugita},
\author{K. Ema},
\author{H. Kuwahara}

\address{Department of physics, Sophia University, Tokyo 102-8554, Japan}

\begin{abstract}
Photoinduced effects in a single crystal of bilayered manganites, La$_{2-2x}$Sr$_{1+2x}$Mn$_2$O$_7$ ($x$=0.38), were investigated in a wide range of temperatures by pump-probe measurement at a photon energy of 1.6eV\@.  In a ferromagnetic metallic state, significant enhancement of positive rise in differential reflectivity with a slow relaxing time of hundred picoseconds was observed just below $T_{\rm{C}}$=127K, indicating that the reflectivity change with the slow relaxation time constant is induced by laser heating.  We have also observed an unconventional fast relaxing component that has a time constant of the order of ten picoseconds.  This fast relaxing component, whose absolute value has an asymmetric peak at $T_{\rm{C}}$, is presumably due to short-range correlation of Jahn-Teller distortion.
\end{abstract}

\end{frontmatter}


Since the discovery of colossal magnetoresistance (CMR), perovskite manganites have been intensively investigated from viewpoints of application such as MR device, and of fundamental science such as spin-charge coupled phenomena.  Extensive investigations have revealed that the origin of CMR cannot be fully described by double-exchange interaction alone \cite{Zener,Anderson,de Gennes} although it partly explains some aspects of CMR in manganites with wide bandwidth.  Competing interactions against double exchange, such as Jahn-Teller interaction or charge/orbital-ordering correlation, also play an important role in some novel properties such as CMR in manganites with narrow bandwidth or reduced dimensionality as presented here \cite{Millis1,Roder,Millis2}.  Recent studies of diffuse x-ray or inelastic neutron scattering have revealed that an additional mechanism for CMR is strong electron-lattice coupling due to Jahn-Teller interaction \cite{Shimomura,Doloc,Argyriou,Kubota1}.  Ultrafast optical measurement using a femtosecond laser pulse is also a promising technique for investigation of such electron-lattice coupling.
    Experiments into photoinduced dynamics using this technique have recently been performed in the ferromagnetic phase \cite{Matsuda,Ogasawara1,Ren,Liu,Averitt,Lobad} of perovskite manganites in order to clarify the relaxation process and the ultrafast photo-control of electronic and magnetic states observed in these systems.  Most of the reported studies of photo-dynamics in manganites are concerned with a long-range ordered state such as ferromagnetic or charge-ordered phase, since even in the long-range ordered state, the optical response is quite different from conventional materials and difficult to interpret.  Advances in the study of the ordered phase of three-dimensional perovskite manganites, e.g., the observation of the demagnetization process through spin-lattice interaction in ferromagnetic phase \cite{Ogasawara1}, gave further insight into the nature of mutual coupling among spin, charge, and lattice degrees of freedom.  By investigating both the paramagnetic and ferromagnetic phases, additional information on the short-range correlation of spin and charge or orbital can be expected.  In order to obtain further insight into competing interactions or short-range correlations, we have investigated photoinduced dynamics using a femtosecond laser pulse in a wide range of temperatures and focused on comparing the behaviors of the paramagnetic and ferromagnetic phases.
  
  Bilayered perovskite manganites La$_{2-2x}$Sr$_{1+2x}$Mn$_2$O$_7$ 0.32$\leqslant$$x$$\leqslant$0.40, which include the $x$=0.38 sample investigated here, are quasi-two-dimensional materials showing CMR\@.  Due to the reduction of dimensionality, competition of the various interactions mentioned above makes its CMR effect much larger than that of the three dimensional manganites.  Bilayered quasi-two-dimensional manganites are, therefore, expected to be suitable for investigating the competition of such interactions.  In this paper, we report on an ultrafast photoexcitation effect over a wide temperature range in the bilayered perovskite manganite La$_{2-2x}$Sr$_{1+2x}$Mn$_2$O$_7$ $x$=0.38, which has the ferromagnetic metallic phase below the ferromagnetic transition temperature $T_{\rm{C}}$=127K and shows the CMR effect ([$\rho$(0)--$\rho$(8T)]/$\rho$(8T)=2700\% at 132K just above $T_{\rm{C}}$.  The observed relaxation process can be described by two typical time constants of the order of ten and one hundred picoseconds, which are called ``fast'' and ``slow'' relaxing components, respectively in this paper.  We estimate each relaxing time constant by using an empirical formula, and discuss the mechanism of each component by using the obtained fitting parameters.

Melt-grown crystal of La$_{2-2x}$Sr$_{1+2x}$Mn$_2$O$_7$ $x$=0.38 was used for the pump-probe measurement.  Stoichiometric mixture of La$_2$O$_3$, SrCO$_3$, and Mn$_3$O$_4$ powder was calcined and sintered.  Subsequently, a single crystal was grown by the floating-zone method at a feeding rate of 9--12mm/h in air.  The sample was characterized by resistivity, powder x-ray diffraction, and magnetization measurements.  The observed ferromagnetic transition temperature $T_{\rm{C}}$ was 127K, which is consistent with the previously reported one \cite{Kubota2}.  From powder x-ray diffraction measurements at room temperature and Rietveld analysis, we confirmed that the grown crystal was a single phase without an impurity phase.  It was of tetragonal symmetry with a space group $I$4/$mmm$ ($a$=0.387142(8)nm, $c$=2.01917(4)nm).  Optical surfaces of $ab$-plane were prepared by cleaving the crystal boule.  The typical size of the prepared sample was 2$\times$2$\times$0.5mm$^3$.  The sample was mounted at the cold head of a temperature-controllable closed cycle refrigerator.  Time-resolved differential reflectivity $\Delta R/R$ at various temperatures was measured using ordinary pump-probe measurements.  Pump and probe pulses were provided by a Ti:sapphire regenerative amplifier system (photon energy for pump and probe pulses 1.6eV, pulse duration 200fs, repetition rate 1kHz).  The time resolution of such measurements is generally limited to the pulsewidth (200fs in our case).  The measurement in our experiment was, however, performed in a 2.5ps interval so that optical responses having timescales shorter than 2.5ps cannot be detected.  The typical power density of the pump pulses was 50$\mu$J/cm$^2$, and the pump and probe pulses were both linearly polarized and orthogonal to each other.  The polarization of the laser was parallel to the pseudo-cubic axis of the cleaved $ab$-plane.

  Figures 1(a) and (b) show the temperature variation of transient differential reflectivity in La$_{2-2x}$Sr$_{1+2x}$Mn$_2$O$_7$ $x$=0.38 below and above $T_{\rm{C}}$, respectively.  Despite the fact that the $\Delta R/R$ curves are seemingly different below and above $T_{\rm{C}}$, each response at different temperatures has a common relaxation process with the following two distinct time constants in all temperature regions.  One is a fast negative component observed as a sharp negative peak in the vicinity of zero time delay, which relaxes within a timescale of ten picoseconds.  The other is a slow component with a relaxation time constant of the order from hundred picoseconds to a few nanoseconds, which appears after the fast component.  The slow component dramatically changes its behavior with temperature, especially near $T_{\rm{C}}$.  
  In order to discuss the above-mentioned changes in signal quantitatively, we have fitted the temporal evolution of the differential reflectivity signals $\Delta R/R(t)$ defined as following empirical formula:
\begin{equation}
\frac{\Delta R}{R}(t)=\left(\frac{\Delta R}{R}\right)_\infty+\left(\frac{\Delta R}{R}\right)_{\rm{fast}} \rm{e}^{-t/\tau _{\rm{fast}}}+\left(\frac{\Delta R}{R}\right)_{\rm{slow}} \rm{e}^{-t/\tau _{\rm{slow}}},
\label{eq:one}
\end{equation}
where ($\Delta R/R)_{\rm{fast}}$ and ($\Delta R/R)_{\rm{slow}}$ are the time-independent amplitudes of the fast and the slow components, respectively, $\tau _{\rm{fast}}$ and $\tau _{\rm{slow}}$ are relaxation times of each component, and $(\Delta R/R)_\infty$ is a time-independent background parameter corresponding to the asymptotic value when the time approaches infinity.  This $(\Delta R/R)_\infty$ represents the offset value from $\Delta R/R$=0, which should recover to the ground state ($\Delta R/R$=0) after a much longer time delay.  Thus, this component can be attributed to equilibrating process - not to the ground state - but rather to a quasiequilibrium one in a nanosecond timescale, which relaxes to the ground state on a much longer timescale.  The observed data were well fitted using the above equation.  We will discuss the temperature dependence of each parameter calculated from the equation in the following paragraphs.

Figure 2 shows the temperature dependence of the parameter $(\Delta R/R)_\infty$ in Eq.(\ref{eq:one}).  As is clearly seen in the figure, the parameter $(\Delta R/R)_\infty$ gradually increases with temperature, showing a sharp peak just below $T_{\rm{C}}$.  Afterwards $(\Delta R/R)_\infty$ abruptly drops, and alters its sign from positive to negative at 140K.  We will discuss $(\Delta R/R)_\infty$ using the parameters of the slow component in Fig.3.  Figures 3(a) and (b) show the temperature dependence of amplitude of the slow component $-(\Delta R/R)_{\rm{slow}}$ and its time constant $\tau _{\rm{slow}}$, respectively.  Below $T_{\rm{C}}$, the absolute amplitude of the a slow component $|(\Delta R/R)_{\rm{slow}}|$ showed a similar temperature dependence to $(\Delta R/R)_\infty$.  The slow component dramatically enhanced just below $T_{\rm{C}}$.  Above 135K, $|(\Delta R/R)_{\rm{slow}}|$ did not show significant structure.  Error bars, as represented in Figs. 3(b) and 4, are too small to be seen in Figs.2 and 3(a).  Similar to the case of temperature dependence of $|(\Delta R/R)_{\rm{slow}}|$, its time constant $\tau _{\rm{slow}}$ also showed peak structure just below $T_{\rm{C}}$ with a local maximum value of 750ps.  After this parameter decreases down to 125ps at 130K, it increases again with increasing temperature as shown in Fig. 3(b).

Let us consider the origin of the time constant $\tau _{\rm{slow}}$ and the temperature variation of $|(\Delta R/R)_{\rm{slow}}|$.  In ordinary metals such as Cu, Ag, and Au, the thermalization time of the electron and lattice system through electron-phonon interaction ranges from 0.5ps to 2ps \cite{Fann,Sun1,Sun2} and a response with a time delay \textgreater100ps after photoexcitation would be attributed to thermal diffusion processes or recombination processes of interband transition leading to the recovery of the ground state  ($\Delta R/R$=0).  In the case observed here, an apparent non-zero (positive) value of $(\Delta R/R)_\infty$ was observed as shown in Fig. 2, which means that this slow relaxing process represents the equilibrating process to quasiequilibrium, i.e. the process that does not recover to the ground state corresponding to $\Delta R/R$=0\@.  Similar slow relaxing behavior to quasiequilibrium has been reported in the ferromagnetic metallic phase of other manganites \cite{Ogasawara1,Ren,Lobad}, and is expected to be a characteristic of these materials.  The inset of Fig. 3(a) shows the temperature derivative of conventional (non-time-resolved) reflectivity $\Delta R/\Delta T$ at the photon energy of 1.6eV, which is the same one as used in our time-resolved optical measurements.  The $\Delta R/\Delta T$ value, an indicator for the reflectivity change with temperature, is calculated using the temperature dependence of conventional reflectivity, which shows a steep increase just below $T_{\rm{C}}$.  Comparing Fig. 3(a) with the inset of Fig. 3(a), both quantities ($|(\Delta R/R)_{\rm{slow}}|$ and $\Delta R/\Delta T$) clearly show a similar temperature dependence.  Judging from the similarity in both quantities, the temperature variation of the slow component ($\Delta R/R)_{\rm{slow}}$ is obviously influenced by the heating effect.  The major action of the pump pulse in a timescale of hundred picoseconds is the transfer of its light energy to the sample as a heat.  In other words, the slow component is a good index for evaluating the heating effect in the sample.  The peak structure in time constant $\tau _{\rm{slow}}$ just below $T_{\rm{C}}$ is also a characteristic of manganites, and can be considered as reflecting the critical slowing down reported in Ref.12.  The observed time constant at the local maximum is 750ps, which is smaller than that in Ref.12.  The reason for the difference between the observed $\tau _{\rm{slow}}$ and the demagnetization time constant in Ref.12 is that the time constant $\tau _{\rm{slow}}$ derived in this paper originates from a heating effect.  In our experiment for transient differential reflectivity, spin-lattice interacting and thermal relaxation processes of the lattice system are indistinguishable in contrast to the case in Ref.12.  For the separation of spin and lattice degrees of freedom experimentally and more accurate discussion about the peak in the temperature dependence of demagnetization time constant, time-resolved magneto-optical Kerr spectroscopy will be necessary.

$(\Delta R/R)_{\rm{slow}}$ shows a small negative value above 140K as well as the case of ($\Delta R/R)_\infty$.  (Note that $-(\Delta R/R)_{\rm{slow}}$ is shown in Fig.3(a).)  The monotonic increase of $\Delta R/R$ toward $\Delta R/R$=0 above 140K (the process occurring without change of the sign of $\Delta R/R$ in contrast to the case of low temperatures) implies that it is due to the relaxation process not to the quasiequilibrium observed at low temperatures below $T_{\rm{C}}$ but to the ground state ($\Delta R/R$=0) within a nanosecond timescale.  This slow relaxation process in the paramagnetic phase above $T_{\rm{C}}$ originates from a thermal process of mainly lattice degree of freedom without spin.   Concerning the time constant $\tau _{\rm{slow}}$ in the corresponding temperature region above $T_{\rm{C}}$, monotonic increase with increasing temperature is observed, as shown in Fig.3(b).  The obtained increase in time constant $\tau _{\rm{slow}}$ with temperature above $T_{\rm{C}}$ cannot be explained by the conventional model at present.  This is because the observed increase in $\tau _{\rm{slow}}$ with increasing temperature showed the opposite tendency to that of the usual thermal-activation process, in which the decay time constant decreases with increasing temperature.  One of the possible reasons for this anomalous behavior is a transition (``crossover'')from the low time constant toward the quasiequilibrating state to large one toward the ground state.  In order to discuss a component that relaxes to the ground state more accurately, we must also take $(\Delta R/R)_\infty$ into account.  $(\Delta R/R)_\infty$ changes its sign from positive to negative at 140K and does not show any significant temperature dependence above 150K\@.  The ratio $|(\Delta R/R)_\infty$/$(\Delta R/R)_{\rm{slow}}|$ is not so small (0.3$-$0.7) that $(\Delta R/R)_\infty$ can not be negligible.  If the temperature dependence of $\tau _{\rm{slow}}$ above $T_{\rm{C}}$ were due to crossover between the large and small characteristic time constants, relaxation time would show saturation to the large one to the ground state and $(\Delta R/R)_\infty$ should reach zero at a certain temperature where the component toward the ground state with large time constant only exists.  Unfortunately, we were not able to observe such evidence, because the time constant to the ground state has a much larger timescale (ten nanoseconds or more) so that time delay we can use was too short.  The increase of the error bars of $\tau _{\rm{slow}}$ as the temperature increases supports this conclusion.

Figure 4 shows the temperature dependence of (a) the amplitude $-(\Delta R/R)_{\rm{fast}}$ for the fast component and (b) the relaxation time $\tau _{\rm{fast}}$ of this component.  In Fig. 4(a), the absolute amplitude $|(\Delta R/R)_{\rm{fast}}|$ shows an asymmetric peak structure $at$ $T_{\rm{C}}$: steep increase was observed below $T_{\rm{C}}$, while a more gradual decrease takes place above $T_{\rm{C}}$.  As shown in Fig. 4(b), the time constant $\tau _{\rm{fast}}$ ranges from 20ps to 35ps below $T_{\rm{C}}$ and from 10ps to 15ps above $T_{\rm{C}}$.  The reason for the large error bars in $\tau _{\rm{fast}}$ below $T_{\rm{C}}$ is due to the complex nature of the time-resolved signal: the negative component within the short time delay is overlapped by the positive one for the long time delay.  (See also the raw data in Fig.1(a).)  This fast relaxing component with a relaxation time of ten picoseconds shows the peak $at$ $T_{\rm{C}}$ in contrast to the case of $|(\Delta R/R)_{\rm{slow}}|$, in which its peak was observed $just\ below$ $T_{\rm{C}}$.  First, we discuss the timescale of this component.  In ferromagnetic metallic manganites \cite{Averitt,Lobad}, typical timescales of time-resolved signals reported in published data can be characterized in two timescales. One is of the order of a few picoseconds and is interpreted as a relaxation process through electron-phonon interaction (our experiment was performed in 2.5ps interval so this component cannot be resolved). The second is a timescale of \textgreater hundred picoseconds and is interpreted as a thermal relaxation process (this component, named as a slow component in this paper, has been discussed in previous paragraph by using Fig.3).  The fast signal with a time constant of ten picoseconds observed in our experiment is unconventional and cannot classified under the above-mentioned typical two-timescale regimes.  Our result suggests that there exists a different relaxation mechanism that originates from the complex properties inherent in manganites.  One possibility for a non-radiative relaxation process with a time constant of ten picoseconds is the existence of some kind of trapping state as discussed below.

It should be noted that the amplitude of this fast component $|(\Delta R/R)_{\rm{fast}}|$ showing the asymmetric peak structure at $T_{\rm{C}}$ indicates that the component is somehow related to the magnetic phase transition.  The origin of this fast component is not clear solely from the present experiment.  It will be discussed in the light of the published data and its interpretation.  In one report \cite{Ren}, the authors interpret a similar unconventional fast transient optical response as phase separation.  In the present system, there has been no report on phase separation in the ferromagnetic metallic region, thus it would be unjustified to attribute our fast component to the same origin \cite{Ren}.  A possible origin for this fast component is photoexcitation and a trapping process of carriers to quasistatic Jahn-Teller (JT) distortion \cite{Doloc,Argyriou,Kubota1}, which interacts with localizing carriers.  The photon energy of 1.5eV, which is near our photon energy, is assigned to a $d$-$d$ transition in optical spectral weight \cite{Jung2,Allen,Quijada}.  Although there is some debate on whether this $d$-$d$ transition is an inter- or intra-atomic transition, both interpretations can explain the optical transition at 1.5eV as arising from JT split-levels.  Above $T_{\rm{C}}$, quasistatic JT distortion \cite{Doloc,Argyriou,Kubota1} has been reported on a timescale of about 1ps, which is a comparable or longer than the timescale of our laser pulsewidth (0.2ps).  Our laser pulse will be able to excite and observe the transition between JT split-levels.  Bearing in mind that photoexcitation detects a spatially averaged effect within the beam radius (350$\mu$m), the signal may be reflecting the volume fraction of this quasistatic JT distortion.  The decrease of the fast component below $T_{\rm{C}}$ may be due to the decrease of volume fraction of the local JT distortion, which is consistent with the delocalization of the charge carriers below $T_{\rm{C}}$.  On the other hand, the decrease above $T_{\rm{C}}$ may be reflecting the change in the dynamic nature of the short-range ordering of JT distortion from the quasistatic one to the dynamic one: using a short laser pulsewidth we can detect this timescale crossover.

 In summary, in the ferromagnetic metallic state, we have observed a transient thermal process that showed a positive rise in reflectivity with a time constant of the order of a hundred picoseconds.  This slow component showed a large enhancement $just\ below$ $T_{\rm{C}}$.  This phenomenon is well explained by the heating effect induced by laser pulses.  We also observed a fast relaxing component that shows a maximum amplitude $at$ $T_{\rm{C}}$.  Dynamical Jahn-Teller distortion is likely to be the origin of this fast component. However, further investigation of, for example, chemical pressure effects will be necessary to obtain further information on the role of lattice distortion.  The study on the fast relaxing component provides new insight into the fast-timescale fluctuating nature of manganites.

{\bf Acknowledgement.} This work was supported by a Grant-in-Aid from the Ministry of Education, Culture, Sports, Science and Technology of JapaniNo.15540351).

\newpage


\newpage

\begin{figure}[htbp]
\begin{center}
\includegraphics[width=1\linewidth]{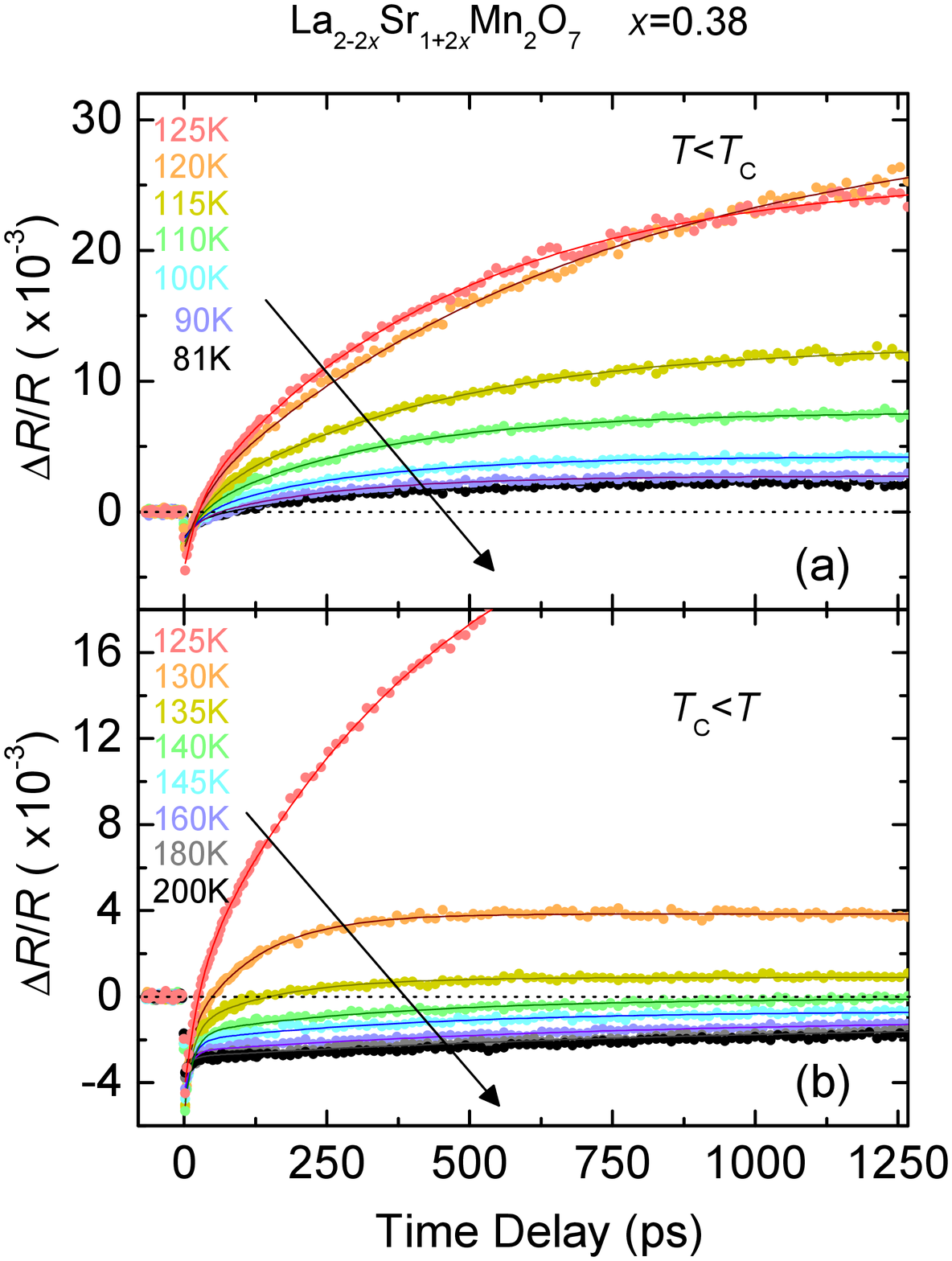}
\end{center}
\caption{Temporal evolution of differential reflectivity $\Delta R/R$ (a) below $T_{\rm{C}}$ (=127K) and (b) above $T_{\rm{C}}$ (and 125K) in La$_{2-2x}$Sr$_{1+2x}$Mn$_2$O$_7$  ($x$=0.38).  Solid lines indicate fitting curves Eq.(\ref{eq:one}).  Horizontal dotted line corresponds to $\Delta R/R$=0.}
\label{Fig.1}
\end{figure} 

\begin{figure}[htbp]
\begin{center}
\includegraphics[width=1\linewidth]{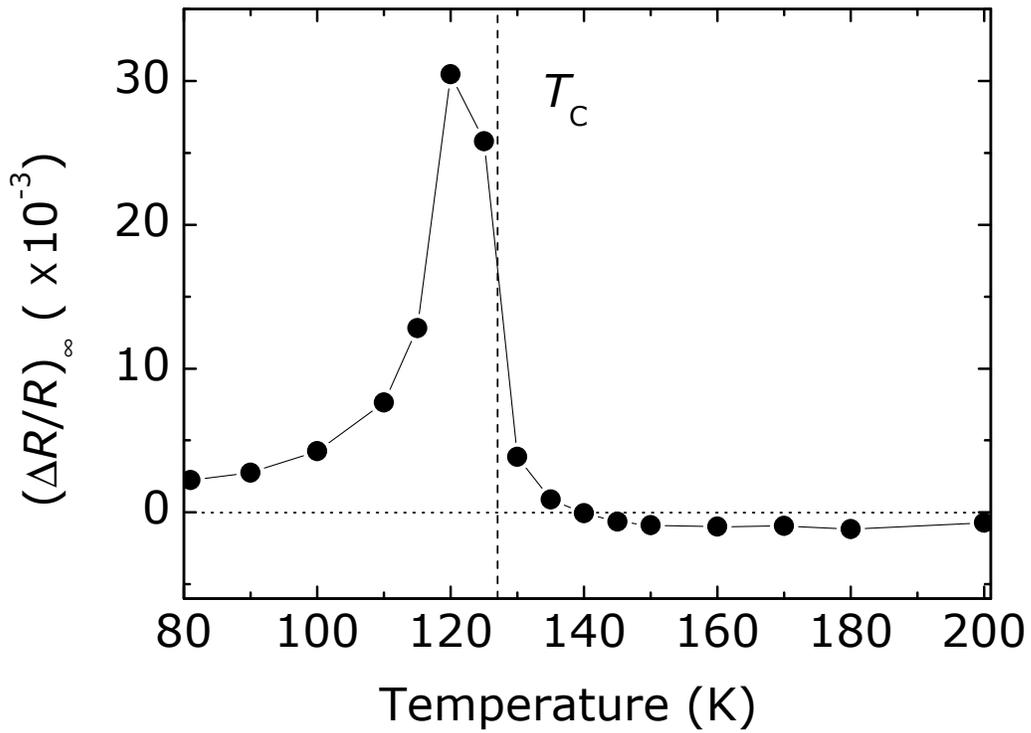}
\end{center}
\caption{Temperature dependence of fitting parameter $(\Delta R/R)_\infty$ in Eq.(\ref{eq:one}).  Solid line is a guide to the eye.  Horizontal dotted lines indicate $(\Delta R/R)_\infty$=0.  Vertical broken line corresponds to $T_{\rm{C}}$=127K.}
\label{Fig.2}
\end{figure} 

\begin{figure}[htbp]
\begin{center}
\includegraphics[width=1\linewidth]{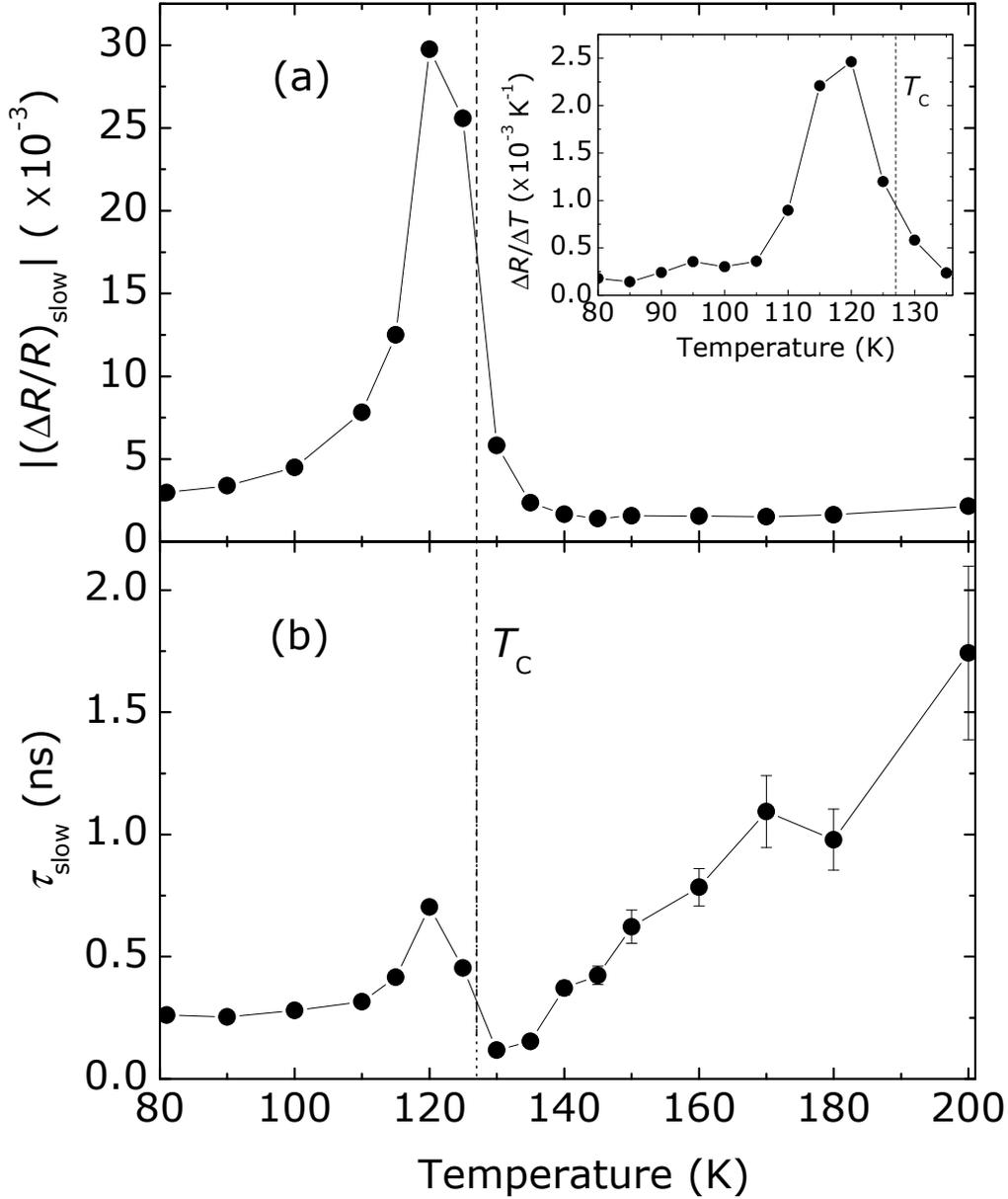}
\end{center}
\caption{Temperature dependence of fitting parameters of Eq.(\ref{eq:one}): (a) amplitude of slow component $-(\Delta R/R)_{\rm{slow}}$, and (b) time constant $\tau _{\rm{slow}}$ for slow component.  Inset in (a) is temperature derivative of conventional (non-time-resolved) reflectivity $\Delta R/\Delta T$ at the same photon energy of 1.6eV\@.  Solid lines in (a), (b), and inset of (a) are a guide to the eye.  Vertical broken lines correspond to $T_{\rm{C}}$=127K.}
\label{Fig.3}
\end{figure} 

\begin{figure}[htbp]
\begin{center}
\includegraphics[width=1\linewidth]{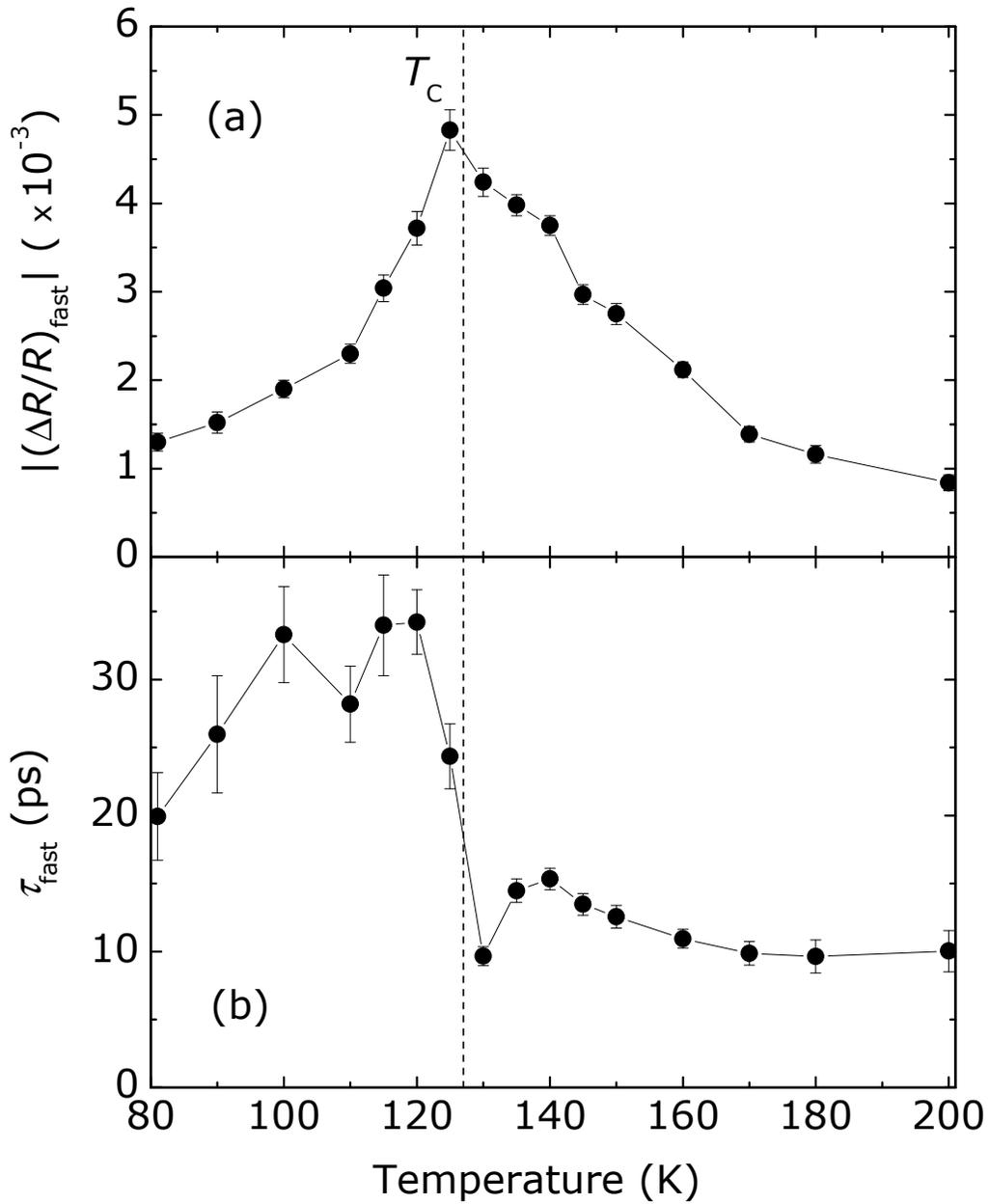}
\end{center}
\caption{Temperature dependence of fitting parameter of Eq.(\ref{eq:one}): (a) amplitude of fast component $-(\Delta R/R)_{\rm{fast}}$ and (b) time constant $\tau _{\rm{fast}}$ for fast component.  Solid lines in (a) and (b) are a guide to the eye.  Vertical broken line corresponds to $T_{\rm{C}}$=127K.}
\label{Fig.4}
\end{figure}

\end{document}